\documentclass[dvips]{article}
\newcommand{\doublespacing}{\let\CS=\@currsize\renewcommand{\baselinesstrech}
{2.0}\tiny\CS}

\begin{document}

\textwidth 16cm
\newcommand{\bd}{\begin{document}}
\newcommand{\ed}{\end{document}}
\newcommand{\bc}{\begin{center}}
\newcommand{\ec}{\end{center}}
\newcommand{\bfr}{\begin{flushright}}
\newcommand{\efr}{\end{flushright}}
\newcommand{\lt}{\left}
\newcommand{\rt}{\right}
\newcommand{\vs}{\vspace}
\newcommand{\hs}{\hspace}
\newcommand{\beq}{\begin{equation}}
\newcommand{\eeq}{\end{equation}}
\newcommand{\lb}{\linebreak}
\newcommand{\pb}{\pagebreak}
\newcommand{\mb}{\makebox}
\newcommand{\fb}{\framebox}
\newcommand{\mc}{\multicolumn}
\newcommand{\ben}{\begin{enumerate}}
\newcommand{\een}{\end{enumerate}}
\newcommand{\bit}{\begin{itemize}}
\newcommand{\eit}{\end{itemize}}
\newcommand{\ol}{\overline}
\newcommand{\un}{\underline}
\newcommand{\lefq}{\lefteqn}
\newcommand{\ba}{\begin{array}}
\newcommand{\ea}{\end{array}}
\newcommand{\beqa}{\begin{eqnarray}}
\newcommand{\eeqa}{\end{eqnarray}}
\newcommand{\beqas}{\begin{eqnarray*}}
\newcommand{\eeqas}{\end{eqnarray*}}
\newcommand{\bfg}{\begin{figure}}
\newcommand{\efg}{\end{figure}}
\newcommand{\bds}{\begin{displaymath}}
\newcommand{\eds}{\end{displaymath}}
\newcommand{\btb}{\begin{tabbing}}
\newcommand{\etb}{\end{tabbing}}
\newcommand{\para}{\parallel}
\newcommand{\pad}{\partial}
\newcommand{\nn}{\nonumber}
\newcommand{\la}{\leftarrow}
\newcommand{\ra}{\rightarrow}
\newcommand{\lgla}{\longleftarrow}
\newcommand{\lgra}{\longrightarrow}
\newcommand{\La}{\Leftarrow}\newcommand{\Ra}{\Rightarrow}
\newcommand{\Lra}{\Leftrightarrow}
\newcommand{\Lgla}{\Longleftarrow}
\newcommand{\Lgra}{\Longrightarrow}
\newcommand{\bm}{\boldmath}
\newcommand{\lan}{\langle}
\newcommand{\ran}{\rangle}
\renewcommand{\a}{\alpha}
\renewcommand{\b}{\beta}
\newcommand{\g}{\gamma}
\newcommand{\G}{\Gamma}
\renewcommand{\d}{\delta}
\newcommand{\eps}{\epsilon}
\newcommand{\Th}{\Theta}
\newcommand{\s}{\sigma}
\newcommand{\lam}{\lambda}
\newcommand{\D}{\Delta}
\newcommand{\vare}{\varepsilon}
\newcommand{\pr}{\prime}
\newcommand{\ro}{\rho}
\newcommand{\nab}{\nabla}
\newcommand{\m}{\mu}
\newcommand{\n}{\nu}
\newcommand{\Sg}{\Sigma}
\newcommand{\p}{\pi}
\newcommand{\R}{I\!\!R}
\newcommand{\om}{\omega}
\newcommand{\Om}{\Omega}
\newcommand{\ze}{\zeta}
\newcommand{\vart}{\vartheta}
\newcommand{\tri}{\triangle}
\newcommand{\f}{\frac}
\newcommand{\iny}{\infty}
\newcommand{\pro}{\propto}
%\input{fqhsc}
%\input{state}
%\input{acc}
%\input{fs}
%\ed
\bc {\huge \bf $(1+1)$ dimensional  Dirac equation with non
Hermitian interaction } \ec

\vs{2cm}

\bc {\large \it A. Sinha{\footnote {e-mail :
anjana23@rediffmail.com}}} \ec

%\vs{.25cm}

\bc {\large and} \ec

%\vs{.25cm}

\bc {\large \it P. Roy{\footnote {e-mail : pinaki@isical.ac.in}}}
\ec

\bc {\large \it Physics \& Applied Mathematics Unit \\
Indian Statistical Institute \\
Kolkata - 700 108} \ec

\vs{1cm}

\vs{1cm}

\bc {\large {\un{Abstract}}} \ec

We study $(1+1)$ dimensional Dirac equation with non Hermitian
interactions, but real energies. In particular, we analyze the
pseudoscalar and scalar interactions in detail, illustrating our
observations with some examples. We also show that the
relevant hidden symmetry of the Dirac equation with such an
interaction is pseudo supersymmetry.

\vs{1cm}

\pb

\section{Introduction}

Non Hermitian Hamiltonians have been studied widely primarily
because of their intrinsic interest \cite{pt} and also for
possible applications \cite{hatano}. One interesting feature of
some non Hermitian interactions, in particular the $\cal{PT}$
symmetric ones is that they have real eigenvalues and such
potentials have been studied in detail within the framework of non
relativistic quantum mechanics. Here our objective is to examine
whether or not $\cal{PT}$ symmetric interactions are possible in
non Hermitian Dirac equation \cite{mudry}. In particular we shall
consider $(1+1)$ dimensional Dirac equation and it will be shown
that $\cal{PT}$ symmetric interactions can be accommodated within
such a framework. It will also be shown that the underlying
symmetry of such a system may be described by {\it pseudo
supersymmetry}.

\section{$(1+1)$ dimensional Dirac equation}

The $(1+1)$-dimensional Dirac equation for a fermion of rest mass
$m$ (in units $\hbar = c = 1 $) is given by \cite{antonio, nogami}
\beq H \psi = E \psi \eeq where \beq H = \alpha p + \beta m +
{\cal{V}} \label{dirach} \eeq $E$ is the energy of the fermion,
$p$ the momentum operator, and $\alpha $ , $ \beta $ are $2 \times
2$ matrices satisfying $ ~ \alpha ^2 ~=~ \beta ^2 ~=~ 1 ~~, ~~
\lt\{ \a, \b \rt\} ~=~ 0. ~~$ A convenient choice of
$\a ~,~ \b$ may be given by the $2 \times 2$ Pauli matrices. \\
The potential matrix ${\cal{V}}$ may be represented as
\cite{antonio} \beq {\cal{V}} = I V_t + \a V_e + \b V_s + \b
\gamma ^5 V_p \label{pot} \eeq where $I$ is the $2 \times 2$ unit
matrix,  $ V_t $ and $V_e$ are the time and space components
respectively, of the $2$-vector potential, and $V_s$ and $V_p$ are
the scalar and pseudoscalar terms. \\ It can be shown easily that
the space component of the vector potential can be gauged away, as
its only contribution is to change the spinors by a local phase
factor. Furthermore, for non Hermitian $V_t $, ${\cal{PT}}$
invariance cannot be accommodated in the $(1+1)$ dimensional Dirac
equation. Consequently, we choose a vanishing time component $V_t
= 0 $, and shall consider only the non Hermitian scalar and
pseudoscalar interactions  in (\ref{pot}), in further detail.

\vs{.5cm}

\noindent {\bf  ${\cal{PT}}$ invariance of the $(1+1)$-dimensional
Dirac equation}

\vs{.5cm}

The operators ${\cal{P}}$ and ${\cal{T}}$ are defined by their
action on the position and momentum operators $x$ and $p$ by
\cite{pt} : \beq \lt. \ba {lcl} {\cal{P}} ~ &:&
~ x \rightarrow -x, \qquad p \rightarrow -p  \\
{\cal{T}} ~ &:& ~ x \rightarrow x, \qquad p \rightarrow -p, \qquad
i \rightarrow -i  \ea \rt\} \label{pt} \eeq For the $(1+1)$
dimensional Dirac Hamiltonian with non Hermitian interactions, to
be invariant under the combined action of ${\cal{PT}}$, i.e., \beq
H \ {\cal{PT}}  = {\cal{PT }} \ H \label{hpt} \eeq one needs to
put a restriction on the choice of $\a$ and $\b$. Since $p$ and
$m$ remain invariant under this transformation, \beq  p  \
{\cal{PT}} = {\cal{PT}} \ p
 \ \ , \qquad \qquad  m \ {\cal{PT}}  =  {\cal{PT}} \ m
\eeq  so should $\a$ and $\b$ : \beq  \a \ {\cal{PT}}  =
{\cal{PT}}  \a  \ \ , \qquad \qquad  \b  \ {\cal{PT}}  =
{\cal{PT}} \ \b  \eeq Hence, to study the $(1+1)$  dimensional
Dirac equation in the framework of ${\cal{PT}}$ symmetric quantum
mechanics, only those non Hermitian interactions can be considered
in (\ref{pot}), which behave under ${\cal{PT}}$ as follows : $V_s
$ should remain invariant while $V_p$ should change sign, i.e.,
\beq V_p ^{{\cal{PT}}} = - V_p \ \ , \qquad \qquad  V_s
^{{\cal{PT}}} =  V_s \eeq

\subsection{Non Hermitian Pseudoscalar Interaction}

First of all, we consider the interaction to be pseudoscalar. \beq
{\cal{V}} (x) = \b \gamma ^5 V_p (x) \eeq A convenient choice of $
\a , \b $ may be \cite{antonio}
\begin{equation} \lt. \ba {lcl}
\displaystyle \a &=& \s _1 = \left( \begin{array}{cc} 0 & 1 \\
 1 & 0 \end{array}  \right) \\ \\
\displaystyle \b &=& \s _3 = \left( \begin{array}{cc} 1 & 0 \\
 0 & -1 \end{array}  \right) \\ \\
\displaystyle \b \gamma ^5 &=& \s _2 = \left( \begin{array}{cc} 0
& -i \\ i & 0 \end{array}  \right) \ea \rt\} \label{ab}
\end{equation}
so that the $(1+1)$ dimensional Dirac equation, with such an
interaction term, reduces to  \beq \lt\{ -i \s _1
\partial _x + \s _3 m + \s_2 V_p (x) \rt\} \psi (x) = E \psi
\label{1dirac} \eeq where
$ \psi (x) $ is a $2$-component spinor : \begin{equation}
\displaystyle \psi = \left( \begin{array}{c} \phi ^{(1)} \\
 \phi ^{(2)} \end{array}  \right)
\end{equation}
Writing (\ref{1dirac}) explicitly, \beq \displaystyle \left(
\begin{array}{cc} \displaystyle m & -
i \partial _x - i V_p \\
 -i \partial _x + i V_p  & - m  \end{array}  \right)
 \left( \begin{array}{c} \phi ^{(1)} \\
 \phi ^{(2)} \end{array}  \right) = \displaystyle \left( \begin{array}{cc} E & 0 \\
 0 & E \end{array}  \right) \left( \begin{array}{c} \phi ^{(1)} \\
 \phi ^{(2)} \end{array}  \right) \eeq where $ \partial _x $
stands for $\displaystyle  \f{d}{dx}$. \\ \\ Under the combined
action of ${\cal{PT}}$, $\s _2$ changes sign while $\s _1$ and $\s
_3$ remain unaltered. Thus the $(1+1)$ dimensional Dirac
Hamiltonian remains invariant under ${\cal{PT}}$ if the
pseudoscalar potential reverses its sign \beq V_p ^* (-x) ~=~ -
V_p (x) \eeq Decomposing the Dirac equation into the upper and
lower components of its spinor, \beq \lt\{ -i \f{d}{dx} - i V_p
\rt\} \phi ^{(2)} = \lt( E - m \rt) \phi ^{(1)} \eeq \beq \lt\{ -i
\f{d}{dx} + i V_p \rt\} \phi ^{(1)} = \lt( E + m \rt) \phi ^{(2)}
\eeq leads to the pair of equations \beq {\cal{H}} _i \phi ^{(i)}
= \lt\{ -  \f{d^2}{dx^2} + U_i (x) \rt\} \phi ^{(i)} = \varepsilon
\phi ^{(i)} \ \ , \ \ \ \ \ \ i = 1,2 \eeq where \beq \varepsilon
=  E^2 - m^2  \label{eps} \eeq \beq U_i =  \lt( V_p ^2 \pm V_p
^{\pr} \rt) \ \ , \ \ \ \ \ \ i = 1,2 \eeq In conventional quantum
mechanics, ${\cal{H}}_{1,2}$ are a supersymmetric (SUSY) pair of
Hamiltonians, with $ V_p (x) $ as the superpotential. The
eigenfunctions of ${\cal{H}}_1$ (${\cal{H}}_2$) are related to
positive (negative) energies of the Dirac Hamiltonian $H$.
Although $U_1 (x)$ and $U_2 (x) $ are different, all eigenvalues,
with the possible exception of the ground state, are shared by $
{\cal{H}}_1$ and $ {\cal{H}} _2 $, and $ \varepsilon \geq 0 $
\cite{nogami}. Defining an intertwining operator $L$ by \beq L  =
 \lt\{ - \frac{d}{dx} + V_p (x) \rt\} \eeq so that
its adjoint is \beq L^{\dagger}  =  \lt\{ \frac{d}{dx} + V_p (x)
\rt\} \eeq the SUSY partner Hamiltonians can be expressed as \beq
{\cal{H}}_1 = L L^{\dagger} \ , \qquad \qquad \qquad {\cal{H}}_2 =
L^{\dagger} L \label{hh} \eeq  The intertwining operators $L$ and
$L^{\dagger}$ generate the supercharges $Q$ and $Q^{\dagger}$
given by \beq
Q = \left(\ba{cc}0 & L \\
0 & 0 \ea \right)~~,~~ \qquad \qquad Q^{\dagger}  ~=~ \left(\ba{cc}0 & 0\\
L^{\dagger} & 0 \ea \right) \eeq   which, in turn, obey the
following closed algebra : \beq \lt\{ Q, Q^{\dagger} \rt\} = H ^2
\ , \qquad \qquad \lt[ Q, H ^2 \rt] = \lt[ Q^{\dagger}, H ^2 \rt]
= 0 \label{hq} \eeq We try to establish a similar hidden symmetry
relationship between the potentials $U_i (x) $ when the pair of
Hamiltonians ${\cal{H}}_1$ , ${\cal{H}}_2$ are not Hermitian,
rather they are pseudo Hermitian with respect to a linear,
invertible, positive definite Hermitian operator $\eta $, i.e.,
\cite{mostafa}, \beq {\cal{H}}_i ^{\dagger} = \eta {\cal{H}}_i
\eta ^{-1} \ \ , \qquad \qquad i=1,2 \eeq (we note here that for
${\cal{PT}}$ symmetric potentials $U_i$, $ \eta $ may simply be
taken as the parity operator ${\cal{P}}$). \\
One can write the partner Hamiltonians ${\cal{H}}_1$ and
${\cal{H}}_2$ in terms of two intertwining differential operators
$L$ and $M$, %
\beq L =  \lt\{ \frac{d}{dx} + V_p (x) \rt\} \eeq %
\beq M  =  \lt\{ - \frac{d}{dx} + V_p (x) \rt\}  \eeq %
as %
\beq {\cal{H}}_1 = L M \ , \qquad \qquad \qquad {\cal{H}}_2 = M
L \eeq %
so that \beq {\cal{H}}_2 M = M {\cal{H}}_1 \ \ ,\qquad
\qquad L {\cal{H}}_2 = {\cal{H}}_1 L \eeq %
Evidently, if $ \phi ^{(1)}  $ is an eigenfunction of
${\cal{H}}_1$ with energy eigenvalue ${\varepsilon} $ , i.e., %
\beq {\cal{H}}_1 \phi ^{(1)} = {\varepsilon} \phi ^{(1)}  \eeq %
then %
\beq \phi ^{(2)} = \displaystyle \f{i}{E+m} M \phi ^{(1)}
\label{efn} \eeq %
is an eigenfunction of ${\cal{H}}_2$ with the
same eigenvalue
${\varepsilon} $, except for the lowest state : %
\beq {\cal{H}}_2 \phi ^{(2)} = \displaystyle \f{i}{E+m}
{\cal{H}}_2 M \phi ^{(1)} = \displaystyle \f{i}{E+m}(M L ) M \phi
^{(1)}  = \displaystyle \f{i}{E+m} M ({\cal{H}}_1 \phi ^{(1)} ) =
{\varepsilon} \lt( \displaystyle \f{i}{E+m} M \phi ^{(1)}  \rt) =
\varepsilon \phi ^{(2)} \eeq Thus $L$ and $M$ intertwine the
Hamiltonians ${\cal{H}}_1 $ and ${\cal{H}}_2 $ in such a way that
$M$ maps the eigenfunctions of ${\cal{H}}_1$ to those of
${\cal{H}}_2$, and $L$ does its converse. It is worth noting here
that $L$ and $M$ are no longer mutually adjoint operators ( $ L
\neq M ^{\dagger} $ ). On the contrary, they are mutually pseudo
adjoint \cite{mostafa, jpa} \beq M ~=~ L^{\#} ~=~ \eta ^{-1}
L^{\dagger} \eta \ \ , \qquad \qquad L ~=~ M^{\#} ~=~ \eta ^{-1}
M^{\dagger} \eta \label{pseudo} \eeq Though ${\cal{H}}_1$ and
${\cal{H}}_2$ being given by (\ref{hh}) are still isospectral,
with the possible exception of the ground state,  it can be shown
by straightforward algebra that (\ref{hq}) no longer holds in such
a situation. Instead, it is replaced by \beq H ^2 =
\{Q,Q^{\#}\} = \left(\ba{cc} {\cal{H}}_1 & 0 \\
0 & {\cal{H}}_2 \ea \right)~  = \left(\ba{cc} L L^{\#} & 0 \\
0 & L^{\#} L \ea \right)~~, \qquad [Q,H ^2 ] = [Q^{\#}, H ^2 ] = 0
~~\label{algebra} \eeq where the pseudo supercharges $Q$ and
$Q^{\#}$ are generated from the intertwining operators $L$ and
$L^{\#}$ as \beq
Q = \left(\ba{cc}0 & L \\
0 & 0 \ea \right)~~,~~ \qquad Q^\# = \eta^{-1} Q^\dagger \eta ~=~
\left(\ba{cc} 0 & 0 \\ L^{\#} & 0 \ea \right) \eeq Thus, in terms
of its components, the pair of Hamiltonians in $(1+1)$ dimensional
Dirac equation with non Hermitian pseudoscalar interaction,
possesses a hidden pseudo supersymmetry.

 We shall illustrate these results by a couple of
explicit examples which have been ${\cal{PT}}$ symmetrized in two
different ways, viz., the ${\cal{PT}}$ invariant Scarf II
potential \cite{jpa, scarf} and the ${\cal{PT}}$ symmetric
oscillator \cite{znojil}.

\vs{1cm}

\noindent {\bf \un{Explicit Examples} :}

\vs{.5cm}

\noindent {\bf (i) ~~ ${\cal{PT}}$ invariant Scarf II potential}

\vs{.5cm}

We consider the following non Hermitian form of $V_p (x) $ \beq
V_p (x) = \lt( p + q \rt) tanh ~ x - i \lt( p - q \rt) sech ~x
\eeq Evidently, \beq V_p ^* (-x) = - V_p (x) \eeq The pseudo
supersymmetric partners $U_i (x) $ reduce to the form  \beq \ba
{lcl} U_1 (x) &=& \displaystyle  \lt\{ V_p ^2 (x) + V_p ^{\pr} (x)
\rt\} \\ \\ &=& \displaystyle \lt[ - \lt\{ 2(p^2+q^2) -(p+q) \rt\}
sech ^2 x - i
(p-q) \lt\{ 2(p+q) -1 \rt\} sech ~x ~~ tanh ~x + (p+q) ^2 \rt] \\
\ea \eeq %
\beq \ba {lcl} U_2 (x) &=& \displaystyle  \lt\{ V_p ^2 (x) - V_p
^{\pr} (x) \rt\} \\ \\ &=& \displaystyle  \lt[ - \lt\{ 2(p^2+q^2)
+(p+q) \rt\} sech ^2 x - i (p-q)
\lt\{ 2(p+q) +1 \rt\} sech ~x ~~ tanh ~x + (p+q) ^2 \rt] \\
\ea \eeq %
and thus are ${\cal{PT}}$ invariant : \beq U_i ^* (-x) = U_i (x)
\eeq In this particular case both the partners belong to the class
of Scarf II potential. The energy eigenvalues and eigenfunctions
of this exactly solvable model are well known \cite{jpa, scarf}.
${\cal{H}}_1$ and ${\cal{H}}_2$ have respective bound state
energies \beq \lt. \ba {lcl} \varepsilon ^{(1)} _n &=&
\displaystyle  \lt\{ 2 \lt( p+q \rt) n - n^2 \rt\} \ , \qquad
\qquad n=0,1,2, \cdots \\ \\ \varepsilon ^{(2)} _n &=&
\displaystyle  \lt\{ 2 \lt( p+q \rt) n - n^2 \rt\}  \ , \qquad
\qquad n=1,2, \cdots \ea \rt\} \eeq with corresponding
eigenfunctions \beq  \ba {lcl} \phi ^{(1)} _n (x) &=&
\displaystyle \f{ \Gamma \lt( n - 2p + \f{1}{2} \rt) }{ n! \Gamma
\lt( \f{1}{2} - 2p \rt)}~ z^{-p}~(z^*)^{-q} ~ P_n ^{-2p -
\f{1}{2}, ~ -2q - \f{1}{2}} (
i~sinh ~x) \ , \qquad  n = 0,1,2, \cdots \\ \\
\phi ^{(2)}_n (x) &=&  \displaystyle \f{i}{E+m} \ \ M \ \phi
^{(1)} _n (x)  \ , \qquad   n=1,2, \cdots \ea  \label{wavescarf}
\eeq where $ \ \ z = \displaystyle \f{1-i~sinh~x}{2} \ \ $ and
$P_n ^{\ \alpha, \beta}$ are the Jacobi polynomials \cite{hand}.
Thus, using (\ref{eps}), the discrete spectrum for the Dirac
Hamiltonian consists of a positive series \beq E_n ^+ ~=~ + \sqrt{
m^2 + \lt\{ 2 \lt( p+q \rt) n - n^2 \rt\} } \qquad \qquad n=0,1,2,
\cdots \eeq and a negative series \beq E_n ^- ~=~ - E_{n+1} ^+ ~=~
- \sqrt{ m^2 + \lt\{ 2 \lt( p+q \rt) (n+1) - (n+1) ^2 \rt\} }
\qquad \qquad n=0,1,2, \cdots \eeq

\vs{.5cm}

\noindent{\bf (ii) ~~ ${\cal{PT}}$ symmetric oscillator}

\vs{.5cm}

Next we move on to the pseudoscalar potential given by \beq
\displaystyle V_p (x) = -z + \f{ \lt( -q \a + \f{1}{2} \rt) }{z}
\eeq where \beq \displaystyle z = x - i \eps \eeq The pseudo
supersymmetric partners being given by \beq \ba {lcl} U_1 (x) &=&
\displaystyle  \lt\{ V_p ^2 (x) + V_p ^{\pr} (x) \rt\} \\
\\ &=& \displaystyle z^2 + \f{ \a ^2 - \f{1}{4} }{ z^2 } +
2 q \a - 2 \\ \ea \eeq \beq \ba {lcl} U_2 (x) &=& \displaystyle
\lt\{ V_p ^2 (x) - V_p ^{\pr} (x) \rt\} \\ \\ &=&
\displaystyle z^2 + \f{ \a ^2 - 2q \a + \f{3}{4} }{z^2}  + 2 q \a \\
\ea \eeq are once again ${\cal{PT}}$ symmetric \beq U_i ^* (-x) =
U_i (x) \eeq and belong to the widely studied class of
${\cal{PT}}$ symmetric oscillator \cite{znojil}. The pair of
Hamiltonians ${\cal{H}}_{1,2}$ possesses real energies given  by
\beq \lt. \ba {lcl} \varepsilon_n ^{ (1)}  &=& \displaystyle
 4n \ \ , \qquad \qquad n=0,1,2,\cdots \\ \\
\varepsilon_n ^{ (2)}  &=& \displaystyle  4n  \ \ , \qquad \qquad
n=1,2,\cdots \ea \rt\} \eeq with corresponding eigenfunctions
(double set, owing to the quasi-parity $q = \pm 1$) \beq \lt. \ba
{lcl} \phi ^{(1)} _{nq} (z) &=& N_{nq} e^{ - \f{ z^2}{2} } z^{-q
\a + \f{1}{2} } L_n ^{ (- q \a )} \lt( z^2 \rt) \ \ , \qquad
\qquad  \qquad n = 0,1,2, \cdots \\ \\
\phi ^{(2)}_{nq} (z) &=& \displaystyle  \displaystyle
\displaystyle \f{i}{E+m} ~ M \ \phi ^{(1)} _{nq} (x) =
\displaystyle  \f{i}{E+m} \lt\{ \f{1}{L_n ^{ (- q \a )} \lt( z^2
\rt)} \f{d}{dx} L_n ^{ (- q \a )} \lt( z^2 \rt) \rt\} \phi ^{(1)}
_{nq}  \ \ , \qquad  n=1,2, \cdots \ea \rt\} \eeq where $ L_n ^{(
\sigma )} $ are the associated Laguerre polynomials \cite{hand}.
In a similar manner, the positive and negative series of the
discrete spectrum possessed by the Dirac Hamiltonian turns out to
be \beq \lt. \ba
{lcl} E_n ^{ \ +}  &=& \displaystyle + \ \sqrt{ m^2 + 4n } \ \  \\
\\  E_n ^{ \ -}  &=& \displaystyle - E_{n+1} ^{ \ + }  ~=~
\displaystyle - \ \sqrt{ m^2 + 4(n + 1) }  \ea \rt\} , \qquad
\qquad n=0,1,2,\cdots \eeq

\subsection{Non Hermitian Scalar Interaction}

Now we shall explore the possibility of a non Hermitian scalar
interaction \beq {\cal{V}} (x) = \b V_s (x) \eeq However, since
${\cal{PT}}$ invariance is neither a necessary nor a sufficient
condition for the reality of the spectrum, we can look for a
$(1+1)$ dimensional Dirac equation with non ${\cal{PT}}$
symmetric, non Hermitian interaction, with real energies. For this
purpose, we can suitably choose $\a $, $\b $ as \cite{nogami} \beq
\a = \s _2 ~, \qquad \qquad \qquad \b = \s _1 \eeq so that the
Dirac equation \beq \lt\{ \a p + \b ( m + V_s ) \rt\} \psi  = E
\psi \eeq can be decoupled and reduced to the following pair of
equations \beq \lt. \ba {lcl} \displaystyle
\lt\{ - \f{d}{dx} + m + V_s \rt\} \phi ^{(2)}  &=& E \phi ^{(1)} \\ \\
\displaystyle \lt\{  \f{d}{dx} + m + V_s \rt\} \phi ^{(1)} &=& E
\phi ^{(2)}  \ea \rt\} \eeq giving \beq {\cal{H}}_i \phi ^{(i)} =
\lt\{ -  \f{d^2}{dx ^2} + U_i (x) \rt\} \phi ^{(i)} = \varepsilon
\phi ^{(i)} , \qquad \qquad i=1,2 \eeq with \beq U_i (x) =  \lt\{
\lt( m + V_s \rt) ^2 - m^2 \mp \f{d V_s}{dx} \rt\} \eeq \beq
\varepsilon = E^2 - m^2  \eeq As is obvious from the above
equations, analogous to the non Hermitian pseudoscalar interaction
dealt with in section $2.1$ above, a non Hermitian scalar
interaction can also be studied in the framework of pseudo
supersymmetry for a mass dependent interaction \beq V_s = (  W_s -
m ) \eeq so that \beq {\cal{H}}_i \phi ^{(i)} = \lt\{ - ~
\f{d^2}{dx ^2} + \widehat{U}_i (x) \rt\} \phi ^{(i)} =
\widehat{\varepsilon} \phi ^{(i)} , \qquad \qquad i=1,2 \eeq where
\beq \widehat{U}_i (x) = \displaystyle \lt( W_s ^2 \mp W_s ^{\pr}
\rt) \ \ , \qquad \qquad i=1,2 \eeq and $ \qquad \qquad \qquad
\qquad \qquad \qquad \qquad
\widehat{\varepsilon} = \displaystyle E^2 $. \\  \\
For example, one can choose a scalar interaction of the form \beq
V_s = \lambda tanh ~ x + ia - m \eeq Evidently, only such energy
dependent scalar interactions can be accommodated within the non
Hermitian framework, which admit real energies.

\section{Conclusion}

To conclude, we have studied the $(1+1)$ dimensional solvable
Dirac equation with non Hermitian scalar and pseudoscalar
interactions, possessing real energies. As explicit examples of
the non Hermitian pseudoscalar interaction, we have considered the
Schr\"{o}dinger equivalent of two ${\cal{PT}}$ invariant cases,
viz., the ${\cal{PT}}$ symmetric Scarf II potential and the
${\cal{PT}}$ oscillator. Additionally, we observed that the
relevant hidden symmetry of the $(1+1)$ dimensional Dirac equation
with non Hermitian interaction is {\it pseudo supersymmetry},
which is in contrast to the supersymmetry of its conventional
Hermitian counterpart.

\pb

\end{document}